\documentclass[aps,prl,twocolumn,showpacs]{revtex4}
\usepackage{amsmath}
\usepackage{graphicx,epsfig,psfrag}
\usepackage{amssymb}
\newcommand{\OLD}[1]{{\bf OLD RM}}

\renewcommand{\Re}{\mathbf{Re}\,}
\renewcommand{\Im}{\mathbf{Im}\,}
\renewcommand{\S}{{\mathbf S}}
\newcommand{\Sa}[1]{\S^\alpha_{#1}}
\newcommand{\w}{\omega}

\begin{document}
\title{Transport in Almost Integrable Models: Perturbed Heisenberg Chains}
\author{P. Jung, R. W. Helmes, and  A. Rosch}
\affiliation{Institute for Theoretical Physics, University of Cologne, 50937
Cologne, Germany}
\date{\today} 

\begin{abstract}
The heat conductivity $\kappa(T)$ of integrable models, like the
one-dimensional spin-$1/2$ nearest-neighbor Heisenberg model, is infinite
%
even at finite temperatures as a consequence of the conservation laws
associated with integrability.  Small perturbations
lead to finite but large transport coefficients
which we calculate perturbatively using exact diagonalization and moment expansions. We show that there are two
different classes of perturbations. 
While an interchain coupling of strength $J_\perp$ leads to $\kappa(T) \propto 1/J_\perp^2$ 
as expected from simple golden-rule arguments, 
we obtain a much larger $\kappa(T)\propto 1/J'^4$ for a weak next-nearest-neighbor interaction $J'$. This can be explained by a new approximate conservation law of the $J$-$J'$ Heisenberg chain.
\end{abstract}
\maketitle
The thermodynamic properties of many experimental systems (like, e.g.,~KCuF$_3$ or MEM-[TCNQ]$_2$) are well
described by a one-dimensional (1D) nearest-neighbor spin-$1/2$ Heisenberg model \cite{experiments}. In such systems, 
 measurements of specific heat or susceptibilities are in quantitative agreement with exact
results derived from the Bethe ansatz. The situation is different when
transport is considered. As is typical for an integrable system \cite{zotos}, the heat
conductivity of the Heisenberg chain is {\em infinite} at any 
temperature \cite{kluemper},
while experimentally  measured transport coefficients are finite. 
In real materials 
the unavoidable presence of (small) perturbations like longer range spin-spin interactions, interchain couplings, disorder or spin-phonon interactions, which break the integrability, are expected to render the heat conductivity finite.
One can therefore expect that 
conductivities at finite temperature $T>0$ are singular functions of terms which
break the integrability. This has to be contrasted with the behavior of thermodynamic quantities and
most other correlation functions which---at least for finite temperatures---vary smoothly as a function of small perturbations (assuming that no
phase transitions are induced). Obviously the 
general
question arises of how transport can
be calculated in ``almost integrable models'', i.e.,  how strongly is the 
%
transport affected by small couplings which break the integrability. 
%

This question
is not only important
for systems well described by integrable Heisenberg or Hubbard models but is
also of relevance for a much broader class of quasi-1D materials. The reason is that
effective low-energy theories in 1D are notoriously integrable. For example, an
arbitrarily complicated two-leg spin-ladder is, at low energies, well described by an
integrable Sine-Gordon model as long as the energy gap $\Delta_E$ is much smaller than
microscopic energy scales like $J$. The term ``well-described'' implies again
that the integrable model can be used for an accurate description of
thermodynamics. To understand transport, however, one has to study again the
effects of {\em small} perturbation (suppressed by powers of $\Delta_E/J$) on
transport. 

%
%
A further reason for our investigations
is the
general theoretical question of how singular are integrable models and how are they
affected by perturbations. While the analog question is well studied in
classical systems with a small number of degrees of freedom, c.f. the famous
Kol'mogorov-Arnol'd-Moser theorem \cite{KAM}, not much is known in many-particle quantum  systems.

Heat transport in spin chains has been subject of intense experimental \cite{sologubenko,hess,kudo} and theoretical \cite{zotos, shimshoni, zotos1, zotosLadder, meisner, prosen}  research  in the recent past. Numerical studies on small systems at high temperatures \cite{zotos1, zotosLadder, meisner} have shown that non-integrable models---in contrast to integrable ones---have  finite transport coefficients. However, the regime of small perturbations, which is probably the most relevant experimentally, is not easily accesible by those methods due to the singular nature of conductivities near the integrable point.  (More results are available for classical systems, see, e.g., \cite{garst, prosen, lepri}). Analytical approaches which calculate transport at low $T$ based on the analysis of slow modes within the memory matrix formalism \cite{RoschPRL, shimshoni, roschFL}
are only valid for systems not too close to an integrable point, as the corresponding slow modes have been neglected. One motivation of the present work was actually the question whether these approximations are valid.

We will consider 1D spin-1/2 models with the Hamiltonian
 \begin{equation}
H=H_0+H_1+H_{1,\perp}
\end{equation}
consisting of an integrable part
\begin{equation}
H_0=J\sum_i \left(S^x_i S^x_{i+1}+S^y_i S^y_{i+1}+\Delta S^z_i S^z_{i+1}\right).
\end{equation}
and a small perturbation $H_1$ or $H_{1,\perp}$ which breaks integrability. We will consider two kinds of perturbations. First, we take into account that in reality next-nearest neighbor spins (nnn) are also weakly coupled. 
\begin{equation}
H_1=J'\sum_i \left(S^x_i S^x_{i+2}+S^y_i S^y_{i+2}+\Delta' S^z_i S^z_{i+2}\right),
\label{nnnpert}
\end{equation}
with  $g'=J'/J\ll 1$. Alternatively, we consider the weak exchange coupling between parallel spin chains. Introducing chain indices $\alpha, \beta$,  we use
\begin{equation}
H_{1,\perp}=J_\perp\sum_{\langle\alpha\beta\rangle}\sum_i \mathbf{S}^\alpha_i\cdot\mathbf{S}^\beta_i,
\label{ladderpert}
\end{equation}
with  $g_\perp=J_\perp/J\ll 1$. 

The heat current operator $J_Q=\sum_i j_i$ is obtained in the usual way from the continuity equation $\partial_t h_i=j_{i}-j_{i+1}$, where $h_i$ is given by $H=\sum_i h_i$. In the following we use a symmetrized version \cite{comment3}:
\begin{eqnarray}
h_{i}&=&\frac{J}{2}\sum_\alpha (\Sa{i-1}\cdot\Sa{i}+\Sa{i}\cdot\Sa{i+1}\nonumber \\ &+& 2 g' \Sa{i-1}\cdot\Sa{i+1}+ 2 g_\perp\sum_{\langle\alpha\beta\rangle}\Sa{i}\cdot\S^\beta_i)\nonumber
\end{eqnarray}
for $\Delta=\Delta'=1$, to obtain for $j_i$,
\begin{multline}
\frac{J^2}{2}\sum_\alpha\Big[2 \Sa{i-1}\cdot(\Sa{i}\times\Sa{i+1}) \Big. +g'(3\Sa{i-2}-4 \Sa{i}+3\Sa{i+2})\cdot \\ (\Sa{i-1}\times\Sa{i+1}) \Big. + g_\perp\sum_{\langle\alpha\beta\rangle}(\Sa{i-1}-\Sa{i+1})\cdot(\Sa{i}\times\S^\beta_i)\Big]+\mathcal{O}(g'^2).\nonumber
\end{multline}

In the integrable case, $H=H_0$, where the heat current is conserved,  $[H, J_Q]=0$, the conductivity $\kappa(\w)$ develops a singular contribution at zero frequency, $\Re\kappa(\w)\propto\delta(\w)$.
This singular behavior suggests considering a perturbation theory for the inverse of $\kappa(\w)$, rather than $\kappa(\w)$ itself. We therefore focus our attention on the scattering rate $\Gamma(\w)$ \cite{comment2} defined by
\begin{equation}
\kappa(\w)=\beta\frac{\chi}{\Gamma(\w)/\chi-i\w}
\label{relkappa}
\end{equation}
where $\chi=\beta \langle \langle J_Q;J_Q \rangle \rangle_{\w=0}$ is the static susceptibility of the heat current. In the integrable case $\Gamma(\w)\equiv 0$, which reproduces the singularity. Accordingly,
 $\Gamma(\w)$ will be small for small $g$ and we can therefore  expand the real part of Eq.~(\ref{relkappa}) in $\Gamma$ at least for finite $\w$,
 \begin{equation}
T \w^2 \Re \kappa(\w)\approx   \Re\Gamma(\w)+\frac{1}{\chi}\Im \frac{\Gamma(\w)^2}{\w+i0}+\dots,
\label{expansion}
\end{equation}
where the left-hand side can be evaluated via the Kubo formula  for small $g$.
To leading order in $g$  we can therefore express 
$\Re\Gamma(\w)=\sum_{n=2}^{\infty}g^n \Gamma_n(\w)$ in terms of a 
correlation function of $\partial_t {J_Q}=i[H,J_Q]$ to get
\begin{equation}
g^2  \Gamma_2(\w)= \frac{1}{\w}
\Re\int_0^\infty dt e^{i(\w+i0) t}\langle [\partial_t J_Q(t),\partial_t J_Q]\rangle_0.
\label{gamma}
\end{equation}
Since in the unperturbed system the heat current is conserved, its time derivative is proportional to the perturbation: $\dot{J}_Q=\mathcal{O}(g)$ and the correlation function (\ref{gamma}) can be evaluated with respect to the unperturbed Heisenberg model.

Equation (\ref{gamma}) is well known in the context of the Mori-Zwanzig memory matrix formalism \cite{forster}. While  (\ref{gamma}) is exact in the limit $g \to 0$ at any {\em finite} frequency, it is important to note that this is not the case for $\w=0$ where the expansion (\ref{expansion}) can become singular. However, in our context it is sufficient to know that Eq.~(\ref{gamma}) gives a rigorous lower bound to the heat conductivity $\kappa(\w=0)$ for small $g$.  This has been shown many years ago for systems which can be described by a Boltzmann equation by Belitz \cite{belitz}. The generalization of this result to almost integrable models will be presented in a forthcoming paper  \cite{inPreparation}. Systematic improvements of
(\ref{gamma}) can also be calculated within the memory matrix formalism  \cite{forster,RoschPRL, shimshoni, roschFL}.

\begin{figure}
\includegraphics[width=\linewidth,clip=]{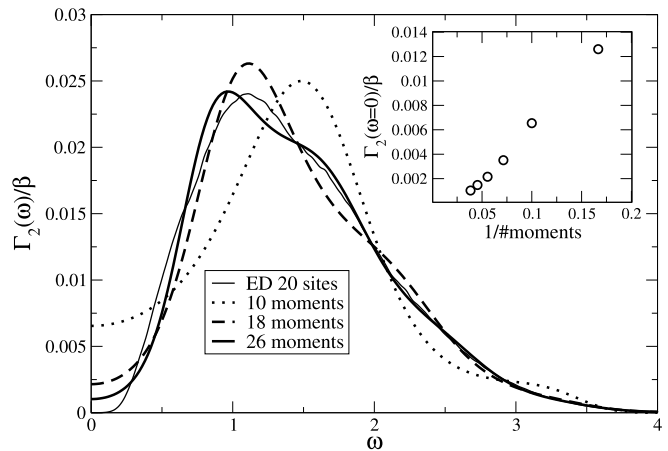}
\caption{Leading order  contribution to the scattering rate $\Gamma(\w)$ from exact diagonalization of a 20-site Heisenberg chain  (thin solid line) with weak nnn coupling $J'$ (at $T=\infty$). To this order the scattering rate vanishes at $\w=0$
, implying anomalous transport
. The same result is obtained  if $\Gamma_2(\w)$ is  reconstructed from $N$ moments ($N=10...26$) using the maximum entropy method. The inset shows that  $\Gamma_2(\w=0)\to 0$ for $N \to \infty$.}
\label{fig1}
\end{figure}
First, we consider the Heisenberg chain with a weak and isotropic ($\Delta'=1$) nnn coupling. 
Figure \ref{fig1} shows the leading order contribution $\Gamma_2(\w)$ to the scattering rate determined from an evaluation of Eq.~(\ref{gamma}) for large $T$ using exact diagonalization. As similar physical quantities (at large $T$) have been reported \cite{zotosNew}
to show surprisingly large finite size effects (not observed in our case) we have also 
reconstructed $\Gamma_2$  from an analytic calculation of its first $26$ moments, $\int_{-\infty}^{\infty} \frac{d \w}{\pi} \w^n \Gamma_2(\w)=\langle[\partial_t^{n-1} J_Q, J_Q]\rangle$, using a high-temperature expansion for an infinite system. 
We have used various methods to obtain  $\Gamma_2(\w=0)$
from these moments  including a continued fraction expansion, the Nickel method \cite{contfracnickel} and the maximum entropy method \cite{maxent} (see  Fig.~\ref{fig1}). Although the curves differ depending on which method is used for reconstruction, all methods consistently show that $\Gamma_2(\w\to 0)$ vanishes. Our exact diagonalization results also show that this is not an artifact of the $T\to \infty$ limit as the limit is smooth (see, e.g.,  Fig.~\ref{fig3}).

We would like to emphasize that the vanishing of the scattering rate $\Gamma(0)$ to lowest order is very surprising both formally and physically. Formally, one would expect that any ``generic'' correlation function of type  (\ref{gamma})  has a finite $\w=0$ limit at any finite temperature. Physically, golden-rule arguments suggest that the breaking of integrability leads to a decay rate of the heat current  of order $J'^2$.
In the following we will first investigate the role of higher order corrections and then 
the influence of other terms which break integrability.  

\begin{figure}
\includegraphics[width=\linewidth,clip=]{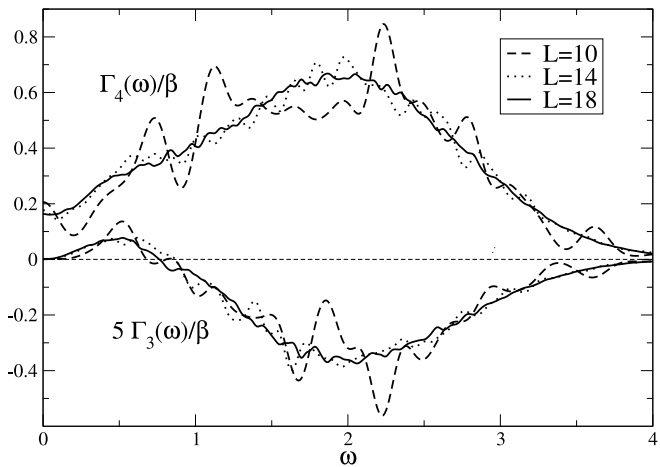}
\caption{Third and fourth order contributions to the scattering rate for various system sizes (see Fig.~\ref{fig1}), the first nonvanishing contribution being of order $g^4$. Note that finite-size effects are small. }
\label{fig2}
\end{figure}
Corrections to $\Gamma$ up to order $J'^4$ are derived starting from Eq.~(\ref{expansion}), where our lowest order result, $\Gamma_2$, is used to determine the term of order $\Gamma^2$. The $\partial_t J_Q$-$\partial_t J_Q$ correlation function is then evaluated to order $J'^3$ and $J'^4$ using the wave functions and energies obtained from the exact diagonalization of $H_0$. The results are shown in Fig.~\ref{fig2}. Since $\Re\Gamma(\w)$ has to be positive and $\Gamma_2(0)=0$, it is not surprising that $\Gamma_3(0)$ also vanishes. $\Gamma_4(0)$, however, is clearly finite.
We therefore conclude that the heat conductivity in the limit $J' \to 0$, $\Delta=\Delta'=1$, has the form
\begin{equation}
\label{xxz}
\kappa \approx  \frac{J^7}{T^2 J'^4  f(T/J)}\approx 
 \frac{0.054(1) \, J^7}{T^2 J'^4} \ \ {\rm for} \ T\to \infty,
\end{equation}
where $f$ is an (unknown) function of $T/J$ only, with $f(x\to\infty)\approx 18.5$ estimated from our exact diagonalization results shown in Fig.~\ref{fig2}.
 Together with the analytical explanation given below this is the main result of our Letter.


We start with the observation that the time derivative of the heat current is linear in $g$ as 
$\left[H_0+g H_1, J_Q \right]=\mathcal{O}(g)$. How can the naive golden-rule argument which suggests a decay rate proportional to $g^2$ fail? This can happen if the presence of slow modes modifies the long-time behavior of the  $\partial_t J_Q$ correlation function as discussed, e.g., in \cite{RoschPRL, shimshoni, roschFL}. We therefore try to construct a new slow mode of the perturbed system $H_0+g H_1$ starting from the conserved heat current $J_0$ of the integrable model $H_0$. Hence, we seek a solution $\tilde{J}_1$ to the equation
\begin{equation}\label{comm}
[H_0+g H_1, J_0+g \tilde{J}_1]=\mathcal{O}(g^2).
\end{equation}
As $[H_0,J_0]=0$, we have to construct a $\tilde{J}_1$  with
\begin{equation}
[H_0,  \tilde{J}_1]=-[H_1, J_0].
\label{commutationrelation}
\end{equation}
Before constructing $\tilde{J}_1$, we investigate the consequences of its existence for the correlator (\ref{gamma}). With $J_Q=J_0+g J_1$ we find
\begin{eqnarray}
- i\dot{J}_Q&=& [g H_1,J_0]+[H_0,g J_1]+\mathcal O(g^2),\\
&=&g [H_0,J_1-\tilde{J}_1]+ \mathcal O(g^2).
\end{eqnarray}
As a consequence, the leading order contribution $\Gamma_2(\w)$ to the scattering rate---by partial integration---may be written as $\Gamma_2(\w)=\w^2 A(\w)$, where $A(\w)$ is the $(J_1-\tilde{J}_1)$ self correlator in the unperturbed system. We therefore conclude that $\kappa(\w=0)$ diverges at least as $1/g^4$ {\em if} $\tilde{J}_1$ exists.
This trick of studying ``readjusted'' approximate conservation laws may well be useful for many other systems with slow modes.
%

We
 turn our attention to relation (\ref{commutationrelation}). To find a solution $\tilde{J}_1$ we make the most general ansatz for it. $\tilde{J}_1$ is  a translationally invariant operator of finite range consisting of a linear combination of products of spin operators. By inserting the ansatz into Eq.~(\ref{commutationrelation}) we obtain a system of linear equations for the unknown coefficients. 
This overdetermined system of equations turns out to have a solution in the case of an isotropic ($\Delta'=1$) nnn perturbation of the Heisenberg model with
\begin{eqnarray}\label{j1}
\tilde{J}_1=- g' J^2\sum_i (\S_{i+1}+\S_{i+2})\cdot(\S_i\times\S_{i+3}).
\end{eqnarray}
The explicit construction of $\tilde{J}_1$ proves the absence of a $J'^2$ contribution to the scattering rate as discussed above. Note that it is not possible to construct a $\tilde{J}_1$ such that the commutator in Eq.~(\ref{comm}) is of order $g^3$ rather than $g^2$.

\begin{figure}
\includegraphics[width=\linewidth,,clip=]{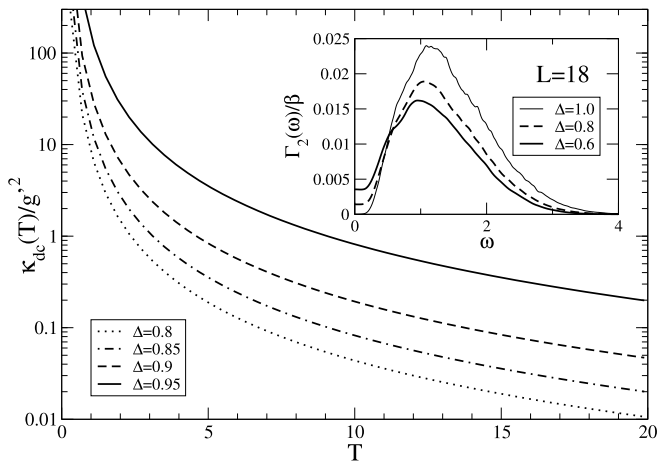}
\caption{Calculated heat conductivity of the anisotropic frustrated chain as a function of temperature for various anisotropies $\Delta=\Delta'$. Inset: leading order contribution to the scattering rate for the isotropic case (thin solid line) as well as with weak anisotropies for $T\rightarrow\infty$.}
\label{fig3}
\end{figure}
While (\ref{j1}) can easily be generalized to the case of an anisotropic XXZ chain with $\Delta\neq 1$,  no solution for $\tilde{J}_1$ exists in the case of an anisotropic nnn {\em perturbation} with $\Delta'\neq 1$. We therefore expect (and confirm numerically) that in the limit of small $J'$ and small but finite $(\Delta'-1)$
\begin{equation}
\label{ xxz}
\kappa \approx  \frac{J^5/T^2}{J'^2 (1-\Delta')^2 h(T/J)}\approx 
 \frac{0.21(2) \, J^5/T^2}{J'^2 (1-\Delta')^2} \ \ {\rm for} \ T\to \infty,
\end{equation}
where $h$ is an (unknown) function of $T/J$ only, the value of which we can determine from the results shown in  Fig.~\ref{fig3} in the limit  $T\to \infty$ .
This figure also shows the $T$ dependence of $\kappa$ for $T\gtrsim J$ where we use Eq.~(\ref{relkappa}) and $\chi$ is calculated to order $g^0$ using exact diagonalization. Large finite size corrections prohibit calculations for $T \ll J$ within exact diagonalization.

\begin{figure}
\includegraphics[width=\linewidth,clip=]{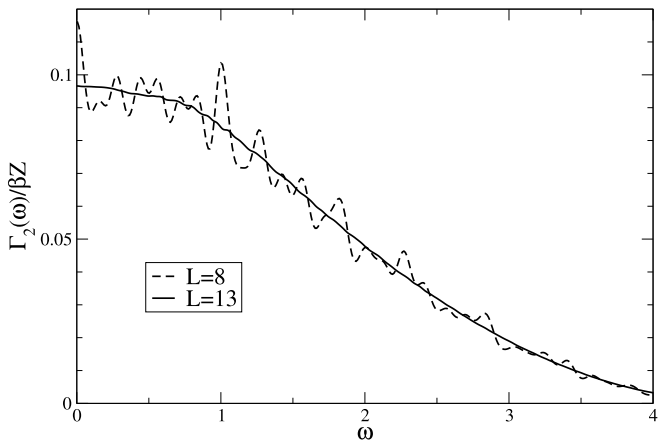}
\caption{Leading order (in $J_\perp/J$) contribution to the scattering rate of weakly coupled spin chains for $T \to \infty$. The finite value at $\w=0$ leads to a conductivity $\kappa\approx 0.091(3) J^5/(Z J_\perp^{2} T^2)$ per chain where $Z$ is the number of nearest-neighbor chains.}
\label{fig4}
\end{figure}
In many experimental systems we expect that the leading term which breaks integrability arises from a weak coupling $J_\perp$ of chains, Eq.~(\ref{ladderpert})
 (or spin-phonon interactions \cite{shimshoni})
.
For this perturbation, Eq.~(\ref{j1}) has no solution and $\kappa\sim 1/J_\perp^2$ can be evaluated at high temperatures from (\ref{gamma}) using exact diagonalization, see Fig.~\ref{fig4}. Our value for the ladder in the limit $J_\perp \to 0$, $\kappa\approx 0.18 J^5/(J_\perp^{2} T^2 )$, seems to be consistent with results of Zotos \cite{zotosLadder} obtained for finite $J_\perp$
using Lanczos diagonalization.

To summarize, we have analyzed the heat transport in spin chains near the integrable point. 
In the presence of a small next-nearest neighbor coupling $J'$, which breaks integrability, one can construct a new approximate conservation law. As a result, the heat conductivity remains extremely high, $\kappa \sim 1/J'^4$. For other perturbations like a weak inter-chain coupling $J_\perp$ this construction is not possible and $\kappa\sim1/J_\perp^2$. 
Thereby we have shown that transport in ``almost integrable models'' depends not only quantitatively, but also qualitatively on the precise way in which integrability is destroyed.
It would be interesting to study experimentally systems in which the strength of $J'$ and $J_\perp$ can be varied systematically, e.g., by chemical substitutions or by pressure.

We thank N.~Andrei, N.~Shah, E.~Shimshoni, and especially X.~Zotos for useful discussions. Part of this work was supported by the German-Israeli Foundation and also the Deutsche Forschungsgemeinschaft through SFB 608.


\begin{thebibliography}{}
\bibitem{comment3}
While the dc heat conductivity is independent of the particular choice for $h_i$ (for a given model),
This is not the case for $\kappa(\w>0)$.
\bibitem{comment2}Strictly speaking, the real scattering rate is $\Gamma(\w)/\chi$.
\end{thebibliography}
\end{document}